\documentclass[twocolumn,11pt,showpacs,preprint,tightenlines,amsmath,amssymb,
groupedaddress]{revtex4}
\usepackage{graphicx,epsfig}
\usepackage{dcolumn}
\usepackage{bm}
\usepackage{amsmath}
\usepackage{psfig}
\def\beq{\begin{equation}}
\def\eeq{\end{equation}}
\def\bea{\begin{eqnarray}}
\def\eea{\end{eqnarray}}
\def\beqa{\begin{equation}\begin{array}{l}}
\def\eeqa{\end{array}\end{equation}}
\def\eqlab#1{\label{eq:#1}}
\def\figlab#1{\label{fig:#1}}
\def\tablab#1{\label{tab:#1}}

\def\eref#1{(\ref{eq:#1})}
\def\Eqref#1{Eq.~(\ref{eq:#1})}

\def\Tabref#1{Table \ref{tab:#1}}

\def\sla#1{#1 \hspace{-2mm} \slash}

\def\barr{\left(\begin{array}{c}}
\def\earr{\end{array}\right)}
\def\bmat{\left(\begin{array}{cc}}
\def\emat{\end{array}\right)}

\def\ga{\gamma} 
 \def\De{\Delta}

\def\mathscr{\mathcal}

\def\3d{3-D}

\def\eqlab#1{\label{eq:#1}}
\def\figlab#1{\label{fig:#1}}
\def\tablab#1{\label{tab:#1}}

\def\eref#1{(\ref{eq:#1})}
\def\Eqref#1{Eq.~(\ref{eq:#1})}

\def\Tabref#1{Table \ref{tab:#1}}

\begin{document}
\title{A dynamical model for pion electroproduction on the nucleon}
\author{George L. Caia}
\email{caia.1@osu.edu} \affiliation{ Institute of Nuclear and
Particle Physics (INPP), Department of Physics and Astronomy, Ohio
University, Athens, OH 45701}
\author{Louis E. Wright}
\email{wright@ohiou.edu} \affiliation{ Institute of Nuclear and
Particle Physics (INPP), Department of Physics and Astronomy, Ohio
University, Athens, OH 45701}
\author{Vladimir Pascalutsa}
\email{vlad@jlab.org}
\affiliation{Department of Physics, College of William \& Mary, Williamsburg, VA 23188 \\
{\em and} Theory Group, JLab, 12000 Jefferson Ave, Newport News, VA
23606}


\begin{abstract}
We develop a Lorenz- and gauge-invariant dynamical model for pion electroproduction in the
resonance region. The model is based on solving of the Salpeter (instantaneous) equation
for the pion-nucleon interaction with a hadron-exchange potential.
We find that the one-particle-exchange kernel of  the Salpeter equation for pion electroproduction
develops an unphysical
singularity for a finite value of $Q^2$. We analyse two methods of dealing with this problem.
Results of our model are compared with recent single-polarization data
for pion electroproduction.
\end{abstract}

\maketitle \section{Introduction}

The electron scattering on the nucleon is the main source of
information about the nucleon structure and the nature of strong
interaction. As such it has been a subject of comprehensive
experimental and theoretical studies for the past several decades.
Most recently, electron and photon beams have  been used at several facilities such as
 JLab \cite{Frolov,Joo}, LEGS
\cite{LEGS}, MAMI \cite{MAMI},  and MIT-Bates \cite{Bates}, to investigate
the electroexcitation of nucleon resonances with an unprecedented precision.
In these works the parameters of the $\De $(1232)-resonance electroexcitation
--- the $\ga^\ast \,N \rightarrow \De $ transition form factors ---
are extracted from the observables of pion electroproduction on the proton
($e \,p\rightarrow e' \,p\,\pi $).

Such determinations of resonance properties from experimental data require
theoretical input,  which, at present,  is provided by the partial-wave analyses
SAID~\cite{SAID} and MAID~\cite{MAID1},
as well as  by various {\it dynamical models}, e.g., ~\cite{Sato2,DMT,Azn:2004}.
We recently developed a new dynamical model of pion electroproduction on the nucleon
\cite{electroCaia}.
The model, henceforth called the OHIO
model, has the advantage of building in the correct
electromagnetic form factors for the nucleon and pion-exchange (Born) contributions.
Therefore we avoid the problem of preserving electromagnetic gauge-invariance
by using the same electromagnetic form factors, as is done in the other descriptions,
see, e.g.,  in~\cite{MAID1,Sato2,DMT}.

Our model is based on solving a quasipotentially-reduced Bethe-Salpeter equation
for pion-nucleon system where the photon is then subsequently attached to describe
the photopion reaction.
We use the {\it equal-time} quasipotential reduction which amounts to neglecting the
relative-energy dependence  of the potential of the Bethe-Salpeter equation thus
leading to the Salpeter equation. In this approximation the pion, nucleon, and resonance
exchanges which take place in the one-hadron-exchange potential appear instantaneously -- the
retardation effects are neglected.

While in the case of pion-nucleon scattering and pion-photoproduction
this quasipotential reduction
 can be implemented rather straightforwardly, inclusion of the virtual photons
technically more difficult because of appearance of new singularities.  These singularities
are associated with the production channels (cuts) that involve particles exchanged in the
driving force. We do not include these production channels in a unitary fashion and
therefore would like to evade corresponding singularities. In the Salpeter formalism this can
be achieved by fixing the relative-energy variable to a $Q^2$-dependent value, where
$Q$ is the photon virtuality. Another viable choice, adopted by us previously \cite{electroCaia, CaiaThesis},
is the ``spectator'' approximation applied to the electroproduction potential.
In this paper we elaborate on the details of our model with an emphasis
on how we deal with the problem of singularities of the Salpeter equation.
We also compare the results of our model with the recent polarization data from JLab.

The paper is organized as follows.
In Sec. II, we describe the model and outline the problem of the exchange singularities.
In Sec. III we present the two choices of quasipotential reduction which evade
these singularities. In Sec. IV we study the model results for both choices,
in comparison with recent experimental data from CLAS at JLab.

\section{The model and particle-exchange singularities}
The OHIO model is based on the unitarity dynamics of the $\pi N$
scattering model presented in \cite{piNscatt}, where it is shown to be possible
to approach the electromagnetic induced reactions in a way which
satisfies the unitarity in the photo-pion channel
space, and hence obeys exactly the Watson theorem.
The model is based on a $\pi N$-$\ga N$ coupled-channel
equation which, when solved to first order in the electromagnetic
coupling $e$, leads to the electroproduction amplitude,
$T_{\pi\ga^\ast}=V_{\pi\ga^\ast}+ T_{\pi\pi} G_\pi
V_{\pi\ga^\ast}$, where  $V_{\pi\ga^\ast}$ is the basic
electroproduction potential, $G_\pi$ is the pion-nucleon
propagator and $T_{\pi \pi}$ is the full $\pi N$ amplitude. Thus,
pion rescattering effects are included as the final state
interaction.

The advantage of
this approximation is that the scattering equation has to be
solved iteratively only for the $\pi N$ scattering amplitude and
then one can evaluate the electromagnetic amplitudes in a one loop
calculation.
\begin{figure}[ht]
\begin{center}
\includegraphics[totalheight=0.2\textheight]{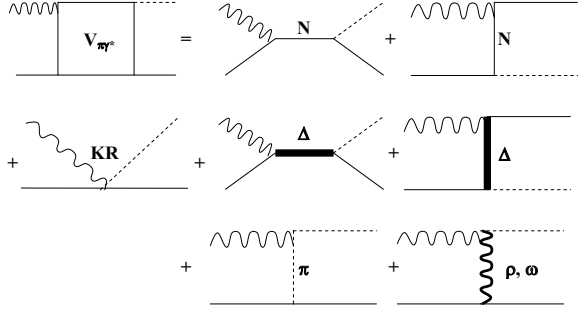}
\caption{\label{photoprod_potential} The Born, vector-meson, and
$\De$-isobar contributions included in the electroproduction
potential.}
\end{center}
\figlab{photoprod_potential}
\end{figure}
Our model for the pion production potential (i.e., the driving
term $V_{\pi \gamma^{*}}$), is given in \cite{VladThesis,
CaiaThesis} and includes the following tree-level contributions:
$N$ direct and crossed terms, $t$- channel $\pi$, $\omega$ and
$\rho$ exchanges, the Kroll-Ruderman (contact) term, and the
direct and crossed $\Delta$ terms (see
Fig.\ref{photoprod_potential}). The model for the $\pi N$ driving
term (i.e. $V_{\pi \pi}$) has been described in detail in
\cite{piNscatt, VladThesis}.

The possibility of a large negative mass squared of the
virtual photon poses a serious problem in the case when one
calculates the off-shell elements of the matrix $V_{\pi \gamma^{*}}$
(i.e. $Q^{2} \ne 0$). Namely, singularities are encountered in the
integration path when the one-loop integration is performed.  In the
following we will show how this happens, and give a prescription for
avoiding this problem. The singularities only  arise in the
$t$- and $u$-channel exchange terms.

Consider for instance the denominator of the nucleon propagator in the case of
the $u-$channel exchange:
\begin{eqnarray}
\eqlab{denuchannel} S_{N}(p-k)=\frac{\sla p -\sla
k+m_{N}}{u(Q^{2},W,|\vec{q}\hspace{1mm}''|,\cos(\theta))-m_{N}^{2}}
\end{eqnarray}
where \begin{eqnarray} \eqlab{umandelstam} &&
u(Q^{2},W,\vec{q}\hspace{1mm}'',\cos \theta))-m_{N}^{2}\nonumber
\\&&=\left(E_{N}-k_{0}\right)^{2}-
\left(\vec{p}-\vec{k}\right)^{2}-m_{N}^{2} \nonumber \\
&& =\left(E_{N}-\beta \sqrt{s}+q''_{0}\right)^{2} \nonumber
\\&&-\left(\sqrt{(\vec{p}+\beta
\vec{P}-\vec{q}\hspace{1mm}'')^{2}+m_{N}^{2}}\right)^{2},
\end{eqnarray}
where $m_{N}$, $m_{\pi}$ are the masses of the nucleon and pion,
respectively,  $W=\sqrt{s}=P_{\mu} \cdot P^{\mu}$ is the total
CM energy of the system, and the relative intermediate energy and
momentum are designated by $q''_o$ and $\vec {q''}$ respectively.
In \Eqref{umandelstam} we have made use
of the fact that the incoming nucleon has 4-momentum defined by
$p^{\mu}=(E_{N},\vec{p})$ (it is fully {\it on-shell}), and the
outgoing $\pi$ has 4-momentum $k^{\mu}=(k_{0},\vec{k})$, which
written in terms of the relative 4-momentum of the outgoing
channel has the energy $k_{0}=\beta \sqrt{s}-q''_{0}$ and the
3-momentum $\vec{k}=\beta \vec{P}-\vec{q}\hspace{1mm}''$, where
the Lorentz scalars $\alpha$ and $\beta$ are: $\alpha=p \cdot
P/s=(s+m_{N}^{2}-m_{\pi}^{2})/2s$ and $\beta=k \cdot
P/s=(s-m_{N}^{2}+m_{\pi}^{2})/2s$. We have also assumed that all
the kinematics are in the CM frame (i.e.,
$P^{\mu}=(P^{0},0)=(\sqrt{s},0)$). In this frame, and from
kinematics considerations the asymptotic energies of the incoming
nucleon and outgoing pion are:
$\omega_{\pi}=\left(W^{2}-m_{N}^{2}+m_{\pi}^{2}\right)/2W, $ and
$E_{N}=\left(W^{2}+m_{N}^{2}+Q^{2}\right)/2W, $ respectively. The
{\it on-shell} 3-momentum of the incoming nucleon is
$|\vec{p}|=\sqrt{E_{N}^{2}-m_{N}^{2}}$.

The problem of singularities arises when the denominator of the
propagator described by the function \bea\eqlab{poleu_ch}
f&=&\left[E_{N}-\omega_{\pi}+q''_{0}-\sqrt{\left(\vec{p}+\vec{q}\hspace{1mm}''\right)^{2}
+m_{N}^{2}}\right]\nonumber \\ &\times
&\left[E_{N}-\omega_{\pi}+q''_{0}
+\sqrt{\left(\vec{p}+\vec{q}\hspace{1mm}''\right)^{2}+m_{N}^{2}}\right],\nonumber\\\eea
vanishes. Note that in the equal time approximation which we use
for the real photon case, $q_{0}''=0$. The relative asymptotic
energy of the nucleon-pion system is given by:\bea
E_{N}-\omega_{\pi}=\frac{2m_{N}^{2}-m_{\pi}^{2}+Q^{2}}{2W},\eea
while the square root term ranges from $m_{N}$ to $\infty$ as the
integration variable $|\vec{q}\hspace{1mm}''|$ ranges over all
possible values. Since the minimum value of $W$ is
$m_{N}+m_{\pi}$, the function $f$ cannot vanish for $Q^{2}=0$.
(Note that it is the first term in $f$ which could possibly have a
{\it zero}, but for $Q^{2}=0$ this term is always negative.)
However, as $Q^{2}$ increases and $E_{N}-\omega_{\pi}$ becomes
larger (by the addition of $Q^{2}/2W$ as compared to the $Q^{2}=0$
case), the first term in $f$ can vanish. It vanishes
at the smallest value of $Q^{2}/2W$ when $\vec{p}$ and
$\vec{q}\hspace{1mm}''$ are antiparallel and equal in magnitude.

\subsection{ $Q^{2}/2W$ choice}

The  singularity in the $u$-channel nucleon exchange can be evaded by fixing the
relative energy $q_{0}''=-Q^{2}/2W$, instead of to zero as in the equal-time reduction.
 This choice has the advantage
that the propagators in the $u$-channel exchanges are not modified at the
photon point. Similar considerations also apply for the case of
the $u-$channel $\Delta$ exchange, $t-$channel $\pi$, $\rho$,
$\omega$ exchanges, as well as for the case of the hadronic form
factors, such as the pion (monopole), and the rho and omega (one
boson) form factors. The same choice for $q_{0}''$ works for the
u-channel $\Delta$ exchange while for the $t$-channel terms and
hadronic form factors, one must choose $q_{0}''=+Q^{2}/2W$.  We
shall refer to these modifications of the propagators and form
factors as the $Q^{2}/2W$ approximation.

The main objection to this choice would be  that it violates current conservation
even at the {\it on-shell} values of the potential matrix ($V_{\pi
\gamma^{*}}$). This problem could be fixed by a global
restoration of the current such as: \beq \eqlab{procedure} J^\mu
\rightarrow {J'}^\mu(Q^2)= J^\mu -n^{\mu}\frac{q \cdot J}{q \cdot
n},\eeq where $J^{\mu}$ is the $4-$component electromagnetic
current and $n^{\mu}$ is an arbitrary $4-$vector. However, since
one of the main goals of our work was to construct a gauge
invariant current, at least for the {\it on-shell} tree level,
such a restoration of current conservation is unsatisfactory.

Instead, we choose to  restore the electromagnetic
gauge-invariance in the following way.
The isospin decomposition of the pion electroproduction
amplitude is done as follows: \bea \eqlab{isospindec} m^{\mu}_{(T)}&=&
\frac{1}{3} \tau_{a} \tau_{3} m^{\mu}_{(1/2)}+\tau_{a}
m^{\mu}_{(0)}\nonumber\\&+&\left ( \delta_{a3}-\frac{1}{3}\tau_{a}
\tau_{3} \right )m^{\mu}_{(3/2)}\eea where $\tau$ are the usual
Pauli matrices, index {\it a}  stands for the pion isospin states
($\pi ^{\pm}$, $\pi ^{0}$) and the lower index, in brackets,
refers to the total isospin.
Using \Eqref{isospindec} one can
determine the contribution of each exchange to the isospin
decomposed amplitudes.
\begin{subequations}\eqlab{isosp_matrix}\bea
m^{\mu}_{(T=\frac{1}{2})} &=& 3m_{(s,N)}^{\mu }-m_{(u,N)}^{\mu
}+4m_{(t, \pi)}^{\mu}\nonumber\\&+&4m_{(KR)}^{\mu }+m_{(t,
\omega)}^{\mu }+2m_{(u, \Delta)}^{\mu}\\
m^{\mu}_{(T=\frac{3}{2})} &=& 2m_{(u,N)}^{\mu
}-2m_{(t,\pi)}^{\mu}-2m_{(KR)}^{\mu }\nonumber\\&+&m_{(t,
\omega)}^{\mu}+\frac{3}{2}m_{(s, \Delta)}^{\mu}+\frac{1}{2}m_{(s, \Delta)}^{\mu }\\
m^{\mu}_{(T=0)} &=& m_{(s,N)}^{\mu }+m_{(u,N)}^{\mu }+m_{(t,
\rho)}^{\mu }.\eea
\end{subequations}
The gauge invariance of the electromagnetic
interactions can be imposed by the {\it current conservation}
condition, $q_{\mu}\cdot m^{\mu}_{T}=0$, for all values of the
isospin: $T=0,$ $\frac{1}{2}$, $\frac{3}{2}$. Due to the violation
of current conservation introduced by our approximation in the
denominators of the {\it on shell} potential matrix $m^{\mu}_{T}$,
Eqs. \eref{isosp_matrix} have to be modified accordingly. In the
following we introduce the correction terms necessary for each
isospin channel at tree level.   The following notation will be
used for  this derivation: the unmodified denominators are denoted
by a unprimed symbol, while the modified denominators, which
include the $''x^{\mu}=(Q^{2}/2W,\vec{0})''$ approximation, are
denoted by a primed symbol. The $u-$ and $t-$channel denominators
and the hadronic form factor of the exchanged pion can be written:
\begin{subequations}\eqlab{ch_denom}\bea \eqlab{uch_denom}
d_{u} &=&p_{u}\cdot p_{u}-m_{N}^{2} \nonumber \\
d_{u'}&=&(p_{u}-x)\cdot (p_{u}-x)-m_{N}^{2},\eea
\bea \eqlab{tch_denom}d_{t} &=&p_{t}\cdot p_{t}-m_{\pi}^{2} \nonumber \\
d_{t'}&=&(p_{t}+x)\cdot (p_{t}+x)-m_{\pi}^{2},\eea  \bea \eqlab{ff_denom}d_{f_{\pi}}
&=&p_{t}\cdot p_{t}-\Lambda_{\pi}^{2} \nonumber \\
d_{f_{\pi}'}&=&(p_{t}+x)\cdot (p_{t}+x)-\Lambda_{\pi}^{2},\eea
\end{subequations} where $p_{u}=(p-k)^{\mu}$ and $p_{t}=(q-k)^{\mu}$. From
Eqs. \eref{ch_denom} the following relationships between
denominators result:
\begin{subequations}\eqlab{denom_relations}\bea
d_{u'}-d_{u}&=&q_{\mu}\cdot q^{\mu}\Phi
\\d_{t'}-d_{t}&=&q_{\mu}\cdot
q^{\mu} \Psi
\\ d_{f_{\pi}'}-d_{f_{\pi}}&=&q_{\mu}\cdot
q^{\mu} \Psi \eea
\end{subequations} where \bea\Phi&=&\left (Q^{2}+4m_{N}^{2}-2m_{\pi}^{2} \right )/\left (4W^{2} \right
), \\
 \Psi&=&\left (Q^{2}+2m_{\pi}^{2}\right )/\left (4W^{2}\right
 ).\eea
 Using \Eqref{denom_relations} the u and t-channels
propagators, and the pion form factor in our approximation write:
\begin{subequations}\bea
S_{N}'(p_{u})&=&S_{N}(p_{u})\left (d_{u}/d_{u'}\right ) \\
S_{\pi}'(p_{t})&=&S_{\pi}(p_{t})\left (d_{t}/d_{t'}\right
)\\f_{\pi}'(p_{t})&=&f_{\pi}(p_{t}) \left (
d_{f_{\pi}}/d_{f_{\pi}'}\right )^{2}.\eea
\end{subequations} Current conservation has to be satisfied by the isospin
projected {\it on shell} amplitudes (Eqs. \eref{isosp_matrix})
separately. Since the resonant contributions as well as rho and
omega exchanges satisfy the condition by their Lorentz structure,
then the problem reduces just to the Born terms (i.e. nucleon
direct and crossed terms, pion and Kroll-Rudermann terms). Using
the modifications introduced in Eqs. \eref{isosp_matrix} and
Eqs. \eref{denom_relations}, and constructing the invariant
$q_{\mu}\cdot m^{\mu}_{T}$ for each isospin separately the
following violating terms result:
\begin{subequations} \eqlab{violationterms}\bea q_{\mu}
\cdot \Delta m^{\mu}_{\frac{1}{2}}&=&\left [3+ \frac{d_{u}}{d_{u'}}
-4\frac{d_{t}}{d_{t'}}\left ( \frac{d_{f_{\pi}}}{d_{f_{\pi}'}}\right
)^{2}\ \right ]\gamma_{5}\sla k\nonumber \\&+&4\left [
\frac{d_{t}}{d_{t'}}\left ( \frac{d_{f_{\pi}}}{d_{f_{\pi}'}}\right
)^{2}-1\right ]\gamma_{5} \sla q \\ q_{\mu} \cdot \Delta
m^{\mu}_{\frac{3}{2}}&=&-2\left [ \frac{d_{u}}{d_{u'}}-
\frac{d_{t}}{d_{t'}}\left ( \frac{d_{f_{\pi}}}{d_{f_{\pi}'}}\right
)^{2}\right]\gamma_{5}\sla k \nonumber \\&-&2\left [
\frac{d_{t}}{d_{t'}}\left ( \frac{d_{f_{\pi}}}{d_{f_{\pi}'}}\right
)^{2}-1\right ]\gamma_{5} \sla q \\ q_{\mu} \cdot \Delta
m^{\mu}_{0}&=&\left (1-\frac{d_{u}}{d_{u'}} \right )\gamma_{5} \sla
k\eea\end{subequations} After some straightforward algebra in
Eqs. \eref{violationterms} the analytical terms necessary for
correcting the violation of the current conservation, introduced by
our approximation, for the {\it on-shell} amplitude are:
\begin{subequations} \eqlab{correctionterms}\bea
\Delta m^{\mu}_{\frac{1}{2}}&=&\frac{\gamma_{5} \sla k\Phi
q^{\mu}}{d_{u'}}+\frac{-4\Psi q^{\mu}\gamma_{5}(\sla k-\sla q)
}{d_{t'}}\Omega\\\Delta m^{\mu}_{\frac{3}{2}}&=&\frac{-2\gamma_{5}
\sla k\Phi q^{\mu}}{d_{u'}}+\frac{2\Psi q^{\mu}\gamma_{5}(\sla
k-\sla q)}{d_{t'}}\Omega\\\Delta m^{\mu}_{0}&=&\frac{-\gamma_{5}
\sla k\Phi q^{\mu}}{d_{u'}}\eea\end{subequations} where \bea
\Omega=1-\frac{(d_{t'}-Q^{2}\Psi)(2d_{f_{\pi}'}+Q^{2}\Psi)}
{d_{f_{\pi}'}^{2}}\eea Comparing the isospin factors from
Eqs. \eref{correctionterms} with those from Eqs. \eref{isosp_matrix}
one can exactly identify where each of the contributions from
Eqs. \eref{correctionterms} has to be incorporated into the
calculations. Notice that, as expected, they do not have an effect
on the transverse components of the current, hence only the
longitude multipoles would be affected by this correction.

While these additional terms clearly restore current conservation at
tree level, when we carry out the full dynamical calculation using
this procedure or the spectator approximation, we also impose
current conservation numerically in the {\it off-shell}
contributions. We confirmed that the use of a numerical restoration
of current conservation at the tree level reproduced the same
results as given by the analytical results given above.

\subsection{Spectator choice}
Previously \cite{electroCaia, CaiaThesis}
we have used a different method of avoiding the particle-exchange singularities
namely the {\it spectator approximation}~\cite{Gross} made in the electroproduction potential.
As pointed out in Refs.~\cite{GrS93,PaT99} the choice of the spectator particle depends on whether
the potential is of the $t$-channel or $u$-channel exchange.
For the $u-$channel exchange
the outgoing pion is the spectator and therefore should be placed on the mass shell. Then the
outgoing nucleon has the energy: \beq \eqlab{u_spectator}
E_{N'}=W-\sqrt{\vec{q}\hspace{1mm}''^{2}+m_{\pi}^{2}}.\eeq  Under
the spectator approximation the first term in \Eqref{poleu_ch}
becomes equal to $W-2\sqrt{\vec{q}\hspace{1mm}''^{2}
+m_{\pi}^{2}}-\sqrt{(\vec{p}+\vec{q}\hspace{1mm}'')^{2}+m_{N}^{2}}$,
which remains negative for all values of
$|\vec{q}\hspace{1mm}''|$.

If the potential is of the form of $t-$channel exchange the
outgoing nucleon is set on its mass shell. Thus the
outgoing pion energy is: \beq \eqlab{t_spectator}
\omega_{\pi}=W-\sqrt{\vec{q}\hspace{1mm}''^{2}+m_{N}^{2}},\eeq
which also avoids the singularities in the $t$-channel terms. An
advantage of the spectator approximation is that the on-shell
potential matrix ($V_{\pi \gamma^{*}}$) does not violate current
conservation, since this approximation, at the tree level,
corresponds to the asymptotic kinematics.

The main disadvantage of the spectator approximation is
that it modifies the ${\it on-shell}$ behavior even at the photon
point, where no anomalous singularities arise in the u-channel and
t-channel terms. In order to describe the $Q^{2}=0$ data in the
spectator approximation, it is necessary to refit some of the
electromagnetic couplings (the most dramatic change arose in the
magnetic coupling of $\Delta$, $G_{M}$). We have done this refitting and all
the results presented in \cite{electroCaia, CaiaThesis} were
calculated using this approximation.

\section{Results and Discussion}Using the $Q^{2}/2W$ approximation with the restoration of
current conservation as described above, we have fitted the major
electroproduction multipoles that have been extracted from
experiment using the coupling constants determined by pion
photoproduction.

\begin{figure}[ht]
\begin{center}
\includegraphics[totalheight=0.17\textheight]{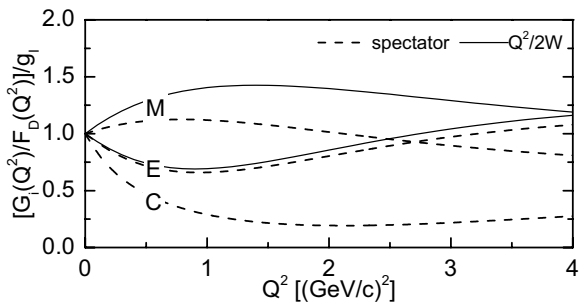}
\caption{\label{deltaff_dipoleff} The $Q^{2}$ dependence of the
determined $N \rightarrow \Delta$ electromagnetic form factors
divided by the dipole form factor, at $W=1.232$ $GeV$. }
\end{center}
\figlab{photprod_potential}
\end{figure}

The parametrization of the $\ga N\De$ form factors we universally
use the form as in \cite{electroCaia}: \beq \eqlab{gg} G_I (Q^2) =
g_{I}\,\frac{1+(Q^{2}/A_I) \, e^{-Q^2/B_I}
}{(1+Q^{2}/\Lambda_I)^2}\,,I=M,\,E,\,C. \eeq Here we have built in a
constraint from  perturbative QCD (pQCD) such that these form
factors fall as $Q^{-4}$ (modulo logs) for asymptotically large $Q$,
see e.g.~\cite{Carlson98}. In \Tabref{NDeltaffparam} are shown the
parameters we used in the $N\leftrightarrow \Delta$ form factors
using the spectator approximation ($II$), while ($I$) denotes the
same parameters using our $Q^{2}/2W-$modification.
\begin{table}[htb]
\begin{center}\vspace{0.35in}
\begin{tabular}{c|c|c|c|c|}
 \quad&$g_{I(II)}$&$\Lambda_{I(II)}$&$A_{I(II)}$&$B_{I(II)}$\\
 \hline
$G_{M}$&2.67(3.10)& 0.71(0.60) & 1.21(1.02) & 1.40(1.20)\\
\hline
 $G_{E}$&0.11(0.05) & 0.50(0.50)  & $-1.10(-1.10)$ & $2.00(2.00)$\\
\hline
$ G_{C}$&-0.38(-0.18) & 0.82(0.78) & $-0.90(-0.90)$ & 1.00(1.00)\\
\end{tabular}
\caption{Parameters used in the $N\leftrightarrow \Delta$ transition
form factors to fit $M_{1+}^{(3/2)}$, $R_{EM}$ and $R_{SM}$. ($I$)
$\pm Q^{2}/2W$ modification in the u- and t-channel terms and ($II$)
spectator approximation.} \tablab{NDeltaffparam}
\end{center}
\end{table}
In Fig. \ref{deltaff_dipoleff} we plot the $Q^{2}$ dependence of
the determined electromagnetic form factors in comparison with the
standard dipole form factor $F_{D}(Q^{2})$ using $\Lambda$=$0.71$
$GeV^2$. The solid lines are the form factors determined in the
$Q^{2}/2W-$approximation, while the dashed lines are the same form
factors determined using the spectator approximation. Note that
all the form factors are normalized to $1$ at the photon point and
that the coupling constants for the two fits are different as
given in Table I.
One sees that the magnetic form factor $G_{M}(Q^{2})$, in both
parameterizations, is overall harder than the dipole at low $Q^{2}$,
but it tends toward the dipole values at higher $Q^{2}$, and the two different methods of
regularizing the scattering equation lead to moderate effects in the $Q^2$ dependence of the form factor.
The electric and Coulomb form factors
$G_{E}(Q^{2})$ and $G_{C}(Q^{2})$ start much softer than the dipole; but
 $G_{E}$  at higher $Q^{2}$ tends to return to
 $F_{D}$, while $G_{C}(Q^{2})$  remains a fraction of $F_{D}$ at high $Q^{2}$.
Unlike the magnetic case, both the electric and Coulomb form factors $G_{E}$ and $G_{C}$ have almost the same $Q^{2}$
dependence in both fits. Note that in all cases the overall trend of
the $N \leftrightarrow \Delta$ transition form factors as a function
of $Q^{2}$ is similar. We can consider the differences as a measure
of the uncertainty of our model.

\begin{figure}[ht]
\begin{center}
\includegraphics[totalheight=0.375\textheight]{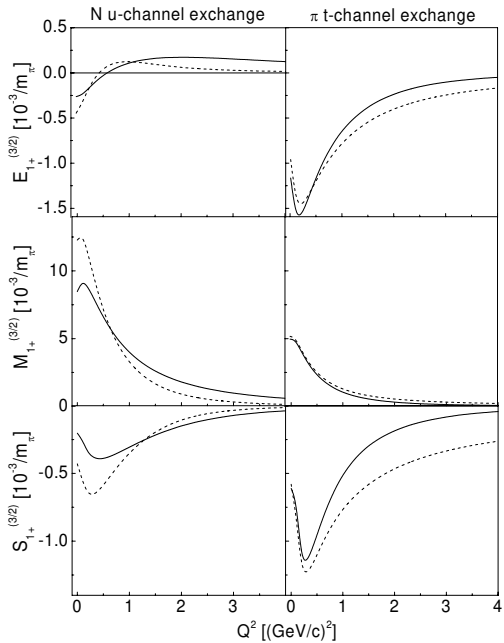}
\caption{\label{contributions_Nu_pit1} Nucleon u-channel (left)
and pion t-channel (right) contributions at $W=1.232$ $GeV$ for
the resonant multipoles. Solid lines are the multipoles determined
using $Q^{2}/2W$ approximations while dashed lines represent the
same calculation using spectator approximation. See text for
explanations.}
\end{center}
\figlab{photprod_potential}
\end{figure}

We have been trying to determine quantitatively the effect of the
two approaches used in avoiding the singularities in the u- and
t-channel terms. Why do the coupling constants have to be modified
from one approach to the other?  The method of avoiding the
singularities affects only the u- and t-channel contributions.
Hence, in Fig. \ref{contributions_Nu_pit1} we plot our calculations
of the resonant multipoles with the full rescattering model, but in
the driving term $V_{\pi \gamma^{*}}$ we include just the pion pole
(right panel) contribution (t-channel pion exchange) or just the
nucleon u-channel exchange (left panel). The t-channel contribution
to the $M_{1+}$ multipole is weakly affected by this modification,
while the u-channel changes significantly. This shift results in
$g_{M}$ changing from 2.67 to 3.10. The contribution to the
$E_{1+}$ multipole at low $Q^{2}$ between model ($I$) and model
($II$) is quite large, resulting in $g_E$ changing from 0.112
to 0.048. The differences in the two models at higher $Q^{2}$ are
not so large since the effects in the t- and u-channel terms
partially cancel. The largest effect is in the $S_{1+}$ multipole.
The u-channel contribution varies significantly at low $Q^{2}$
between models ($I$) and ($II$), while at larger $Q^{2}$ it is the
t-channel contribution which varies. Using the $Q^{2}=0$ results we
had to change $g_{C}$ from -0.38 to -0.18.
Depending on the size of the u- and t-channel contributions to the
various multipoles, the values at $Q^{2}=0$ can change significantly
and the extracted $Q^{2}$ dependence for $G_{M}$, $G_{E}$ and
$G_{C}$ may change from model ($I$) to model ($II$). Fortunately,
the overall $Q^{2}$ dependence as shown in Fig.
\ref{deltaff_dipoleff} is not changed very much. The nucleon
exchange term has a larger effect in the $M_{1+}$ than the pion
pole, in fact, the u-channel change leads to a modification of the
shape of the fitted $G_{M}(Q^{2})$. In the case of the $E_{1+}$
quadrupole, the $Q^{2}$ dependence is not affected in any of the
instances, hence we did not have to refit the $Q^{2}$ dependence of
$G_{E}(Q^{2})$. The $Q^{2}$ dependence of $G_{C}(Q^{2})$ had to be
slightly modified since we see an overall strengthening of both the
pion pole and nucleon crossed terms (above $Q^{2}\approx 1$
$(GeV/c)^{2}$) from model ($II$) to model ($I$), hence we had to
 make $G_{C}(Q^{2})$ slightly harder from model ($II$) to model ($I$). We
did not plot the effects of these two models in the $\Delta$ crossed
term but we analyzed them, and as expected, they did not play an
important role in the resonant multipoles (most of the $\Delta$
contribution to the resonant multipoles comes from the direct and
the Born terms).

\begin{figure}[ht]
\begin{center}
\includegraphics[totalheight=0.375\textheight]{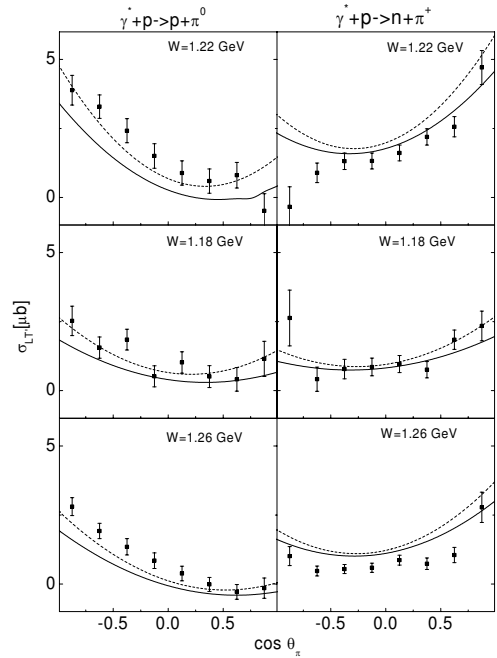}
\caption{\label{sig_LT} $\sigma_{LT'}$ versus $\cos \theta_{\pi}$
for $p(e,e'n) \pi^{+}$ and $p(e,e'p) \pi^{0}$ at $Q^{2}=0.4
(GeV/c)^{2}$. The data points are from \cite{siglt,alt}. The lines
are the same as in Fig.\ref{contributions_Nu_pit1}}
\end{center}
\figlab{sig_LT}
\end{figure}

\begin{figure}[ht]
\begin{center}
\includegraphics[totalheight=0.375\textheight]{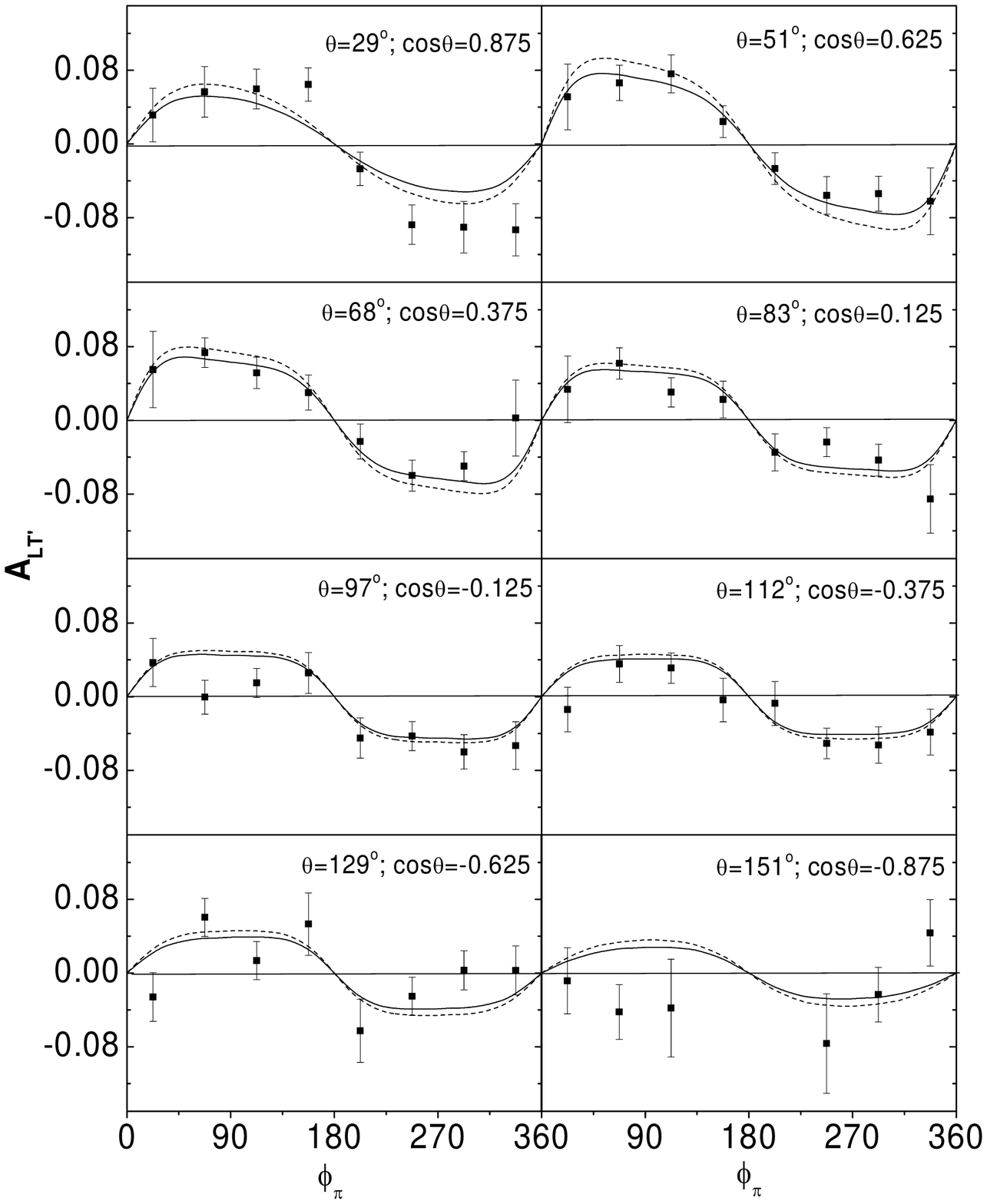}
\caption{\label{A_LT} Beam asymmetry versus $\phi_{\pi}$ for
$p(e,e'n) \pi^{+}$ at $Q^{2}=0.4 (GeV/c)^{2}$, $W=1.22$ $GeV$ and
for fixed $\theta$. The data points are from \cite{alt} and the
lines are the same as in Fig.\ref{contributions_Nu_pit1}}
\end{center}
\figlab{A_LT}
\end{figure}

In Fig. \ref{sig_LT} and Fig.\ref{A_LT} we compare our calculations
with some recent results from CLAS \cite{alt,siglt}. This
asymmetry measurement is important in determining the $t-$channel
pion pole and contact Born terms contributions in $\pi^{+}n$,
which otherwise are weak in the $\pi^{0}p$ channel. Measurements
of beam asymmetry $A_{LT'}$ where made only for the neutral
channel, but since in this reaction the non-resonant amplitude
strongly interferes with the imaginary part of the dominant
$\Delta(1232)$ $M_{1+}^{(3/2)}$ it is difficult to extract the
information about the non-resonant contribution to the
observables. The resulting amplitude for $\pi^{0}p$ is strongly
dependent on the rescattering correction and largely dependent on
the way the model generates the width of the resonance.

The calculated angular distribution of $\sigma_{LT'}$ for the
$\pi^{+}n$ channel show a strong forward peaking for energies
around the $\Delta$, in contrast to $\pi^{0}p$ channel which shows
backward peaking. In Fig.  \ref{sig_LT} and Fig.  \ref{A_LT} we plot
the asymmetry $A_{LT'}$ and the longitudinal-transverse polarized
structure function $\sigma_{LT'}$, respectively. The relationship
between these two quantities, following the reference
\cite{MAID_polarization} conventions, is as follows: \bea
A_{LT'}=\sqrt{2\epsilon_{L}(1-\epsilon)}R_{LT'}\sin \phi
_{\pi}/\sigma_{0},\eea where \bea \sigma_{0}&=&R_{T}+\epsilon_{L}
R_{L}+\sqrt{2\epsilon_{L}(1+\epsilon)}R_{TL}\cos
\phi_{\pi}\nonumber\\&+&\epsilon R_{TT}\cos2\phi_{\pi}, \eea where
$R_{i}$ ($i=T,L,LT,TT$) are the response functions,
$\epsilon=\left (1+2|{\bf q}|^{2}/Q^{2} \tan
^{2}\left(\theta_{e}/2 \right ) \right )^{-1}$ is the degree of
transverse polarization and $\epsilon_{L}=\left (Q^{2}/\omega^{2}
\right )\epsilon$ is the degree of longitudinal polarization of
the virtual photon. Therefore the longitudinal-transverse
polarized structure function shown in Fig. \ref{sig_LT} is:
$\sigma_{LT'}=R_{LT'}/\sin \theta_{\pi}$, which is written in
terms of the electromagnetic multipoles, \bea R_{LT'}=&-&\sin
\theta_{\pi}Im \{(F_{2}^{*}+F_{3}^{*}+\cos
\theta_{\pi}F_{4}^{*})F_{5}\nonumber
\\&+&(F_{1}^{*}+F_{4}^{*}+\cos \theta_{\pi}F_{3}^{*})F_{6} \},\eea
where $F_{i}$ with $i=1,\ldots ,6$, are the CGLN amplitudes
\cite{CGLN} which are defined in the appendix of
\cite{MAID_polarization}.
The two approximations (solid lines is the
$Q^{2}/2W$-approximation and dashed line is the spectator
approximation) are used in our calculations and a fairly good
agreement is obtained in both cases. The largest difference can be
noticed in the case of the asymmetry $A_{LT'}$ for the neutral
channel for c.m. total energy equal to the $\Delta$ mass, while
for the other kinematics there is no practical difference. Since
these observables are directly proportional to the longitudinal
component of the electromagnetic current, then the overall
difference can be a direct measure of the uncertainty of our model
related to the theoretical error introduced by either of the
approximations.

\begin{figure}[ht]
\begin{center}
\includegraphics[totalheight=0.375\textheight]{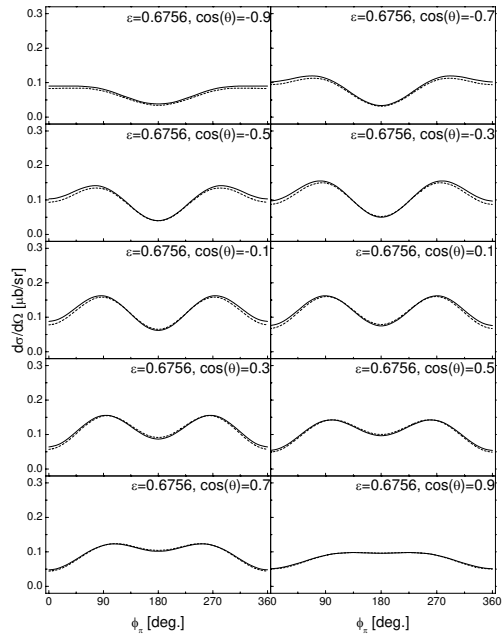}
\caption{\label{Q2_5GeV} Differential cross section versus
$\phi_{\pi}$ for $p(e,e'p) \pi^{0}$ at $Q^{2}=5 (GeV/c)^{2}$,
$W=1.23$ $GeV$ and for fixed $\theta$. The lines are the same as in
Fig.\ref{contributions_Nu_pit1}}
\end{center}
\figlab{Q2_5GeV}
\end{figure}

\begin{figure}[ht]
\begin{center}
\includegraphics[totalheight=0.375\textheight]{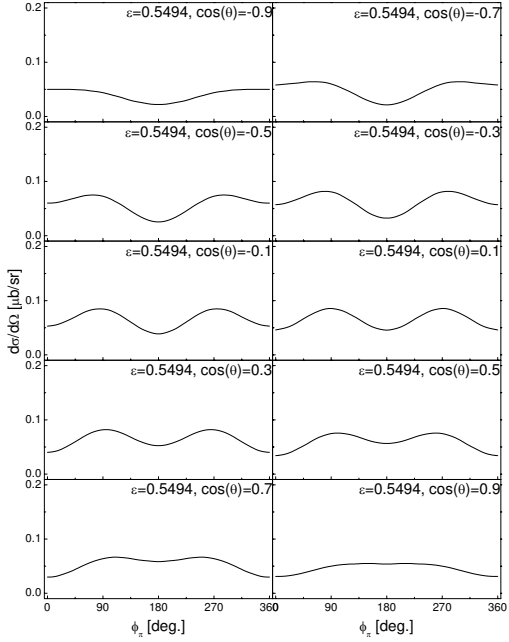}
\caption{\label{Q2_6GeV} Differential cross section versus
$\phi_{\pi}$ for $p(e,e'p) \pi^{0}$ at $Q^{2}=6 (GeV/c)^{2}$,
$W=1.23$ $GeV$ and for fixed $\theta$. Only the $Q^{2}/2W$
approximation is shown, since the spectator gives almost identical
predictions.}
\end{center}
\figlab{Q2_6GeV}
\end{figure}

In view of the upcoming data from JLab for higher values of the
$Q^{2}$, in Figs. \ref{Q2_5GeV} and \ref{Q2_6GeV} we show the
prediction of our model for differential cross sections although we
only fit our $\Delta$ electromagnetic form-factors up to $Q^{2}=4
(GeV/c)^{2}$ (see \cite{electroCaia}). For  the differential cross
section both approximations give very similar results, proving once
again that the cross section is not sensitive enough to possible
theoretical errors in the model.
\section{Conclusions}
A dynamical model for pion electro-production was first introduced
in \cite{electroCaia} where it was used for determining the $\Delta$
electromagnetic form-factors. The major theoretical uncertainty in
our model is the treatment of the u- and t-channel terms in the
potential $V_{\pi \gamma^{*}}$. For large $Q^{2}$ (i.e., large
negative mass squared of the virtual photon), non-physical
singularities occur in these terms when they go off shell in solving
the scattering integral equation.  We have investigated two
approximate ways of solving this problem. The first was the
spectator approximation for these diagrams and the second was to
modify the relative energy in these diagrams by $\pm Q^{2}/2W$. We
have shown that the singularity problem can be partially solved
without affecting the $Q^2=0$ results by proposing an {\it ad-hoc}
solution which avoids these poles. Within the theoretical
uncertainty of how to regularize the u- and t-channel terms for pion
electro-production, this model provides an overall good prediction
of the available data up through the first resonance region.
Comparison with recent single polarization data also shows fairly
good agreement. Furthermore, we predict cross sections at higher
$Q^{2}$ that are being analyzed at Jlab. In our view, the $\pm
Q^{2}/2W$ approximation with restoration of current conservation
provides a good dynamic model for investigating pion
electroproduction at large $Q^2$.  The parameters needed also
describe pion photoproduction quite well and allows the two
processes to be studied in the same model.

\section*{Acknowledgements}
This work was performed in part under the auspices of the U. S.
Department of Energy, under the contracts DE-FG02-93ER40756, DE-FG02-04ER41302,
DE-FG05-88ER40435, and the National Science Foundation under grant
NSF-SGER-0094668. The work of VP is also
supported in part by DOE contract DE-AC05-84ER-40150 under which the Southeastern
Universities Research Association (SURA) operates the Thomas
Jefferson National Accelerator Facility.

\end{document}